\documentclass[11pt,a4paper]{article}

\usepackage{amsmath,amssymb,amsfonts}
\usepackage{physics}
\usepackage{geometry}
\usepackage{hyperref}
\usepackage{graphicx}
\usepackage{cite}

\title{\bf Relativistic Hamiltonian as an emergent structure from information geometry}
\author{Sikarin Yoo-Kong\\
\small The Institute for Fundamental Study(IF), Nareasuan University\\
\small Email: sikariny@nu.ac.th}
\date{}

\begin{document}
\maketitle

\begin{abstract}
We show that the relativistic energy–momentum relation can emerge as an effective ensemble-averaged structure from a multiplicative Hamiltonian when fluctuations of an auxiliary parameter are treated using maximum entropy inference. The resulting probability distribution is uniquely fixed by scale-invariant constraints, which are shown to arise naturally from the Fisher–Rao geometry of the associated statistical manifold. Within this information-geometric framework, the relativistic dispersion relation appears without initially imposing Lorentz symmetry, but as a consequence of statistical averaging and geometric invariance.
\end{abstract}

\section{Introduction}
The Hamiltonian and Lagrangian formalisms are cornerstones of classical and quantum physics, encoding the dynamics and kinematical structure of physical systems. In their conventional additive form, these functionals produce familiar equations of motion and dispersion relations underlying Newtonian and relativistic dynamics. However, non-standard formulations, particularly multiplicative forms of the Lagrangian and Hamiltonian \cite{SSY0}, offer richer perspectives, especially when integrated with statistical or ensemble approaches. A recent study \cite{SSY} demonstrated that treating a family of such multiplicative functionals, parameterised by an auxiliary variable, and averaging over that variable can recover relativistic Lagrangians and Hamiltonians from underlying non-relativistic structures. This suggests two key insights motivating this work: (i) relativistic kinematical features can emerge from statistical averaging over non-relativistic multiplicative Hamiltonians, and (ii) the auxiliary parameter's statistical treatment profoundly shapes the effective dynamics.
\\
\\
From a complementary perspective, information theory and statistical inference provide principled methods for deriving ensemble descriptions from incomplete knowledge. The maximum entropy principle, formulated by Jaynes \cite{Jaynes}, selects the least biased probability distribution that maximises entropy under given constraints, avoiding assumptions beyond available data. This approach has excelled in statistical mechanics, yielding equilibrium distributions from minimal conserved quantities, and extends to broader inference in physics, data analysis, and machine learning \cite{Steve, Pessoa}. Synergistic with this approach is information geometry, which equips parameterised probability families with a natural structure via the Fisher–Rao metric \cite{Amari}. This metric quantifies distribution distinguishability, offering invariant concepts of curvature and coordinates on statistical manifolds, and has illuminated statistical mechanics \cite{Pessoa2}, parameter estimation, and spacetime emergence \cite{Cat}.
\\
\\
This paper combines multiplicative Hamiltonian, entropy maximisation, and information geometry to derive the relativistic dispersion relation as an emergent structure from non-relativistic multiplicative Hamiltonians. We model the auxiliary parameter as fluctuating, inferring its distribution via maximum entropy under scale-invariant constraints. The resulting ensemble-averaged Hamiltonian then exhibits the square-root relativistic form, without presupposing Lorentz symmetry. Notably, these constraints gain geometric meaning: they reflect coordinate and scale information on the statistical manifold, with the Fisher–Rao metric justifying their invariance and the inference process. Thus, relativistic kinematics emerges not as an axiom but as an inference-driven outcome rooted in the manifold's geometry. We shall note here that, in earlier work \cite{SSY}, the relativistic Lagrangian was also shown to emerge from ensemble averaging over a family of multiplicative Lagrangians. However, in the present work, we deliberately adopt a Hamiltonian-first perspective and show that the relativistic square-root Hamiltonian can directly emerge.
\\
\\
This paper is organised in the following manner. In section 2, we introduce the class of multiplicative Hamiltonians and briefly derive the relativistic square-root energy after ensemble averaging. Section 3 formulates the maximum entropy inference of the fluctuating parameter under appropriate constraints. Section 4 develops the information-geometric interpretation by constructing the Fisher–Rao metric on the parameter space and identifying its invariant properties. In section 5, we revisit the entropy constraints in light of the statistical manifold structure and show how they reflect intrinsic geometry. Finally, section 6 concludes with a summary and outlook for future work.



\section{Emergent relativistic Hamiltonian}
\begin{figure}[h]
\centering
\includegraphics[width=1\linewidth]{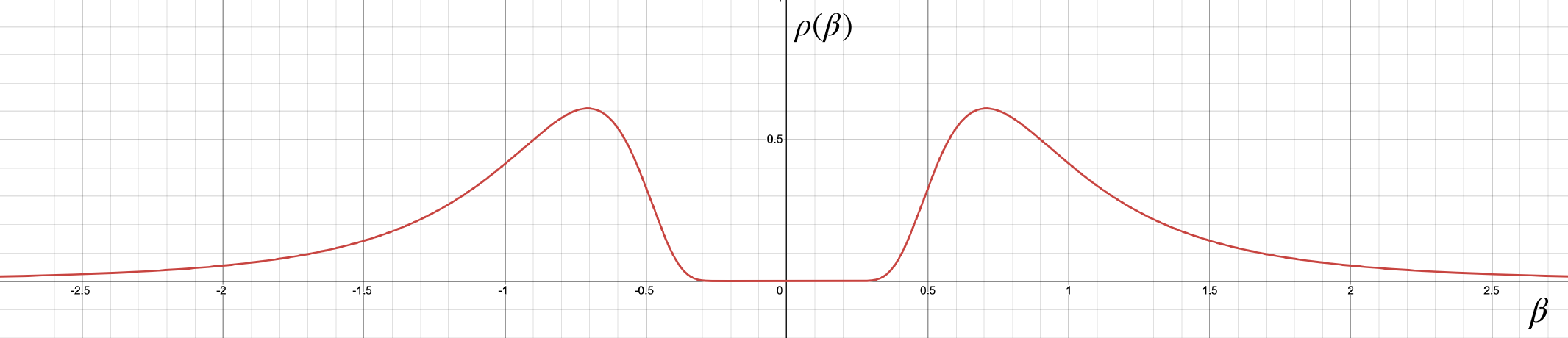}
\caption{\label{dis} The distribution shape of $\rho(\beta)$.}
\end{figure}

In this section, we demonstrate that the relativistic Hamiltonian emerges naturally as an ensemble-averaged quantity when a multiplicative Hamiltonian is combined with fluctuations of an auxiliary parameter $\beta$. 
We begin with the $\beta$-dependent multiplicative Hamiltonian \cite{SSY}
\begin{equation}\label{HB}
H_\beta(p) = m\lambda^2 \beta^2
e^{\frac{p_N^2}{2m\lambda^2\beta^2}}\;,
\end{equation}
where $p_N=m\dot x$ is the spatial momentum and $\lambda$ is a constant with dimensions of velocity. For fixed $\beta$, this Hamiltonian does not exhibit relativistic dispersion and its exponential dependence on $p_N^2$ instead reflects a multiplicative deformation of the standard quadratic kinetic term.
\\
\\
We now assume that the parameter $\beta$ is not fixed, but fluctuates due to incomplete information about an underlying physical or effective scale. The statistical state of the system is therefore described by a probability density $\rho(\beta)$ defined on $\beta \in(-\infty,\infty)$. The ensemble-averaged Hamiltonian is defined as
\begin{equation}
\langle H \rangle
= \int_{-\infty}^\infty d\beta \, \rho(\beta) H_\beta \;.
\end{equation}
We take the distribution as
\begin{equation}\label{rho}
    \rho(\beta)=\frac{2}{\sqrt{\pi}}\frac{e^{-\frac{1}{\beta^2}}}{\beta^4}\;
\end{equation}
with $\int_{-\infty}^{+\infty}d\beta \rho(\beta)=1$, see figure \ref{dis}. 
Substituting Eq.~\eqref{rho} together with the condition $p_N^2<2m\lambda^2$, we have
\begin{equation}\label{relH}
\langle H \rangle
= \frac{m2\lambda^2}{\sqrt{\pi}}
\int_{-\infty}^\infty d\beta \frac{e^{-\frac{1}{\beta^2}+\frac{p_N^2}{2m\lambda^2\beta^2}}}{\beta^2}=\frac{2m\lambda^2}{\sqrt{1-\frac{p_N^2}{2m\lambda^2}}}.
\end{equation}
Equation \eqref{relH} is in the square-root structure. And if we pick $2\lambda^2=c^2$, where $c$ is the speed of light, the average Hamiltonian 
$\langle H \rangle=\gamma mc^2$, where $\gamma=1/\sqrt{1-\dot x^2/c^2}$, is nothing but the relativistic Hamiltonian for a massive free particle. Alternatively, if we introduce the relativistic momentum $p=\gamma m\dot x$, we find that $\gamma^2=1+p^2/m^2c^2$ and $\langle H \rangle=\sqrt{p^2c^2+m^2c^4}$.
\\
\\
Now, we shall apply the Legendre transformation $\langle L\rangle=\langle \dot xp-H\rangle=\langle \dot x p\rangle-\langle H\rangle$. Using $\langle \dot xp\rangle=\dot xp$ and 
\begin{equation}
   \dot x=\frac{\partial \langle H\rangle}{\partial p} =\frac{pc^2}{\sqrt{p^2c^2+m^2c^4}}\;,
\end{equation}
we obtain
\begin{equation}
    \langle L \rangle=-mc^2\sqrt{1-\frac{\dot x^2}{c^2}}\;,
\end{equation}
which is the relativistic Lagrangian \cite{SSY}.
\\
\\
\emph{Remarks}: The square-root form is absent in individual $H_\beta(p)$ and appears only after averaging. No relativistic principles are assumed a priori. The dispersion arises from multiplicative structure and statistical fluctuations. The square-root anticipates a geometric origin, developed below.






\section{The principle of maximum entropy}\label{EP}
The nontrivial $\rho(\beta)$ in section 2 requires justification. With no detailed microscopic origin for $\beta$-fluctuations, we apply Jaynes' maximum entropy principle to select the least-biased distribution consistent with macroscopic constraints. The $\beta$ is an auxiliary parameter and we assign no specific physical meaning, defining its role operationally via inference and geometry.
\\
\\
The Shannon entropy associated with this distribution is
\begin{equation}
S[\rho] = -\int_{-\infty}^\infty d\beta \, \rho(\beta)\ln\rho(\beta).
\end{equation}
The entropy is maximised subject to the following constraints.
\paragraph{The first constraint:}
\begin{equation}
\int_{-\infty}^\infty d\beta \, \rho(\beta) = 1.
\end{equation}
\paragraph{The second constraint:}
\begin{equation}
\left\langle \frac{1}{\beta^2} \right\rangle
= \int_{-\infty}^\infty d\beta \, \frac{\rho(\beta)}{\beta^2} = C_1\;,
\end{equation}
which fixes the average scale of fluctuations in the parametrisation. In the next section, this quantity is directly related to the information-theoretic resolution of the statistical manifold associated with the family of Hamiltonians.
\paragraph{The third constraint:}
\begin{equation}
\left\langle \ln \beta \right\rangle
= \int_{-\infty}^\infty d\beta \, \rho(\beta)\ln \beta = C_2.
\end{equation}
This constraint is invariant under rescaling of $\beta$ and therefore encodes information about the typical location of the statistical state in a scale-invariant manner.
\\
\\
To maximise the entropy subject to these constraints, we introduce Lagrange multipliers $\alpha$, $\gamma$ and $\nu$ and consider the functional
\begin{eqnarray}
\Phi[\rho]
&=&
- \int_{-\infty}^\infty d\beta \, \rho(\beta)\ln\rho(\beta)
- \alpha\!\left(\int_{-\infty}^\infty d\beta \, \rho(\beta)-1\right)\nonumber\\
&&- \gamma\!\left(\int_{-\infty}^\infty d\beta \, \frac{\rho(\beta)}{\beta^2}-C_1\right)
-\nu\!\left( \int_{-\infty}^\infty d\beta \, \rho(\beta)\ln \beta - C_2\right).
\end{eqnarray}
Taking the functional derivative with respect to $\rho(\beta)$ and setting it to zero
\begin{equation}
\frac{\delta \Phi}{\delta\rho(\beta)} = 0,
\end{equation}
gives
\begin{equation}
-\ln\rho(\beta)-1-\alpha-\frac{\gamma}{\beta^2}-\nu \ln\beta=0.
\end{equation}
Exponentiating, the maximum-entropy distribution takes the form
\begin{equation}
\label{rho_raw}
\rho(\beta) =  \frac{C_3}{\beta^{\nu}}e^{-\frac{\gamma}{\beta^2}},
\end{equation}
where $C_3 = e^{-1-\alpha}$ is a normalisation constant. It is not difficult to see that if we pick $\nu=4$, $\gamma=1$ and $C_3=2/\sqrt \pi$, all constraints simultaneously satisfy. 
\\
\\
\emph{Remarks}: The distribution is selected uniquely by entropy maximization under scale-invariant constraints. The $\beta^{-4}$ prefactor (from the $\ln\beta$ constraint) is crucial for the relativistic form after averaging. Though formal here, the constraints' geometric basis is clarified in section 4 via the Fisher–Rao metric.

\section{Statistical manifold and Fisher-Rao metric}\label{SFR}
The emergence of the relativistic Hamiltonian in section 2 suggests that the parameter $\beta$, originally introduced as a fluctuating quantity, plays a deeper structural role. In this section, we show that this role can be made precise by viewing the family of multiplicative Hamiltonians $H_\beta(p_N)$ as defining a statistical manifold equipped with the Fisher–Rao information metric. This geometric viewpoint not only clarifies the origin of the entropy constraints introduced in section 3, but also reveals why the resulting averaged Hamiltonian naturally exhibits a square-root structure.

\subsection{Fisher-Rao metric}
We consider the one-parameter family of multiplicative Hamiltonians
\begin{equation}
    \mathcal{H}=\{H_\beta(p_N)|\beta \in \mathbb{R}\}\;,
\end{equation}
where $\beta$ labels different parametrisations of the same underlying dynamical system rather than distinct physical theories. Accordingly, we interpret $\beta$ as a coordinate on a parameter space $\mathcal M=\{\beta\in \mathbb R\}$, which we endow with a statistical structure.
\\
\\
The construction below should be understood as a formal information-geometric analysis of a parameterized family of positive functions rather than a literal probabilistic model on phase space. 
We first formally associate to each $H_\beta(p_N)$ a positive weight
\begin{equation}
    P(p_N|\beta)=\frac{1}{Z_\beta}H_\beta(p_N)\;,\;\;\;\;Z_\beta=\int dp H_\beta(p_N)\;,
\end{equation}
and regard $\{P(p_N|\beta)\}$ as a one-parameter statistical family. Although the normalisation constant $Z_\beta$ diverges for the present model, this divergence does not affect the Fisher–Rao geometry, which depends only on logarithmic derivatives of $H_\beta$ and is invariant under arbitrary (including divergent) normalisation factors \footnote{For any positive function $f(p,\beta)$ and any positive normalisation factor $A(\beta)$, $g_{\beta\beta}[f]=g_{\beta\beta}[A(\beta)f]$, provided the averages are defined consistently. In this sense, the Fisher–Rao geometry is insensitive to the overall normalisation of the statistical weight, even when the normalisation integral diverges. The information metric thus captures the intrinsic distinguishability of the parametrised family $H_\beta(p_N)$, independent of normalisation artifacts.}.

\subsection{Regulated construction}
To make the above statements explicit, we introduce a momentum cutoff $\Lambda$ and define the regulated normalisation
\begin{equation}
    Z_\beta^{(\Lambda)}=\int_{-\Lambda}^{\Lambda}dp_N H_\beta(p_N)\;,
\end{equation}
together with the regulated probability density
\begin{equation}
    P^{(\Lambda)}(p_N|\beta)=\frac{H_\beta(p_N)}{Z_\beta^{(\Lambda)}}\;.
\end{equation}
The Fisher–Rao metric on $\mathcal M$ is defined by
\begin{equation}
    g_{\beta\beta}^{(\Lambda)}=\int dp_N P^{(\Lambda)}(p_N|\beta)\left(\partial_\beta \ln P^{(\Lambda)}(p_N|\beta)\right)^2\;.
\end{equation}
Using 
\begin{eqnarray}
    \ln P^{(\Lambda)}(p_N|\beta)&=&\ln H_\beta-\ln Z^{(\Lambda)}_\beta\;,
\end{eqnarray}
we find
\begin{eqnarray}
    \partial_\beta\ln P^{(\Lambda)}(p_N|\beta)=\partial_\beta \ln H_\beta(p_N)-\langle \partial_\beta \ln H_\beta \rangle_\Lambda\;,
\end{eqnarray}
so that the metric can be written in the manifestly invariant form
\begin{equation}
    g^{(\Lambda)}_{\beta\beta}=\int_{-\Lambda}^{\Lambda}dp_NP^{(\Lambda)}(p_N|\beta)\left(\partial_\beta \ln H_\beta-\langle \partial_\beta \ln H_\beta \rangle_{\Lambda}\right)^2=\left\langle (\partial_\beta \ln H_\beta-\langle \partial_\beta \ln H_\beta \rangle_{\Lambda})^2\right\rangle_{\Lambda}\;.
\end{equation}
All expectation values appearing in this expression are finite for any finite cutoff $\Lambda$.

\subsection{Scale invariance and logarithmic geometry}
For the multiplicative Hamiltonian $H_\beta(p_N)$, we obtain
\begin{equation}
    \partial_\beta \ln H_\beta =\frac{2}{\beta}-\frac{p_N^2}{m\lambda^2\beta^3}\equiv A\;.
\end{equation}
Introducing the shorthand moments
\begin{equation}
    M_n(\beta)=\langle p_N^{2n}\rangle_\Lambda=\frac{1}{Z_\beta^{(\Lambda)}}\int_{-\Lambda}^{\Lambda} dp_N p_N^{2n}H_\beta(p_N)\;.
\end{equation}
Then, we write
\begin{eqnarray}
    \langle A \rangle_\Lambda&=&\frac{2}{\beta}-\frac{1}{m\lambda^2\beta^3}M_1\;,\\
    \langle A^2 \rangle_\Lambda&=& \frac{4}{\beta^2}-\frac{4}{m\lambda^2\beta^4}M_1+\frac{1}{m^2\lambda^4\beta^6}M_2\;.
\end{eqnarray}
The variance of $A$ must take the form
\begin{equation}
    \text{Var}_\Lambda(A)=\langle A^2\rangle_\Lambda -\langle A \rangle_\Lambda^2=\frac{1}{m^2\lambda^4\beta^6}\left( M_2-M_1^2\right)=\frac{\text{Var}_\Lambda(p_N^2)}{m^2\lambda^4\beta^6}\;.
\end{equation}
Now, we need to compute the variance of $p_N^2$. From the multiplicative Hamiltonian \eqref{HB}, we introduce the dimensionless variable $q=p_N/\beta\sqrt{m\lambda^2}$. Then $p_N^2=m\lambda^2\beta^2 q^2$ and, therefore, $dp_N=\beta\sqrt{m\lambda^2}dq$. The cutoff is also transformed as
\begin{equation}
    p_N=\pm \Lambda \rightarrow q=\pm \frac{\Lambda}{\beta\sqrt{m\lambda^2}}\equiv \pm Q_\Lambda (\beta)\;.
\end{equation}
We find that 
\begin{equation}
 \langle p_N^{2n}\rangle_\Lambda=(m\lambda^2)^{2n}\beta^{2n}C_n(Q_\Lambda)\;,   
\end{equation} 
where 
\begin{equation}
    C_n(Q_\Lambda)=\frac{\int_{-Q_\Lambda}^{Q_\Lambda}dq q^{2n}e^{q^2/2}}{\int_{-Q_\Lambda}^{Q_\Lambda}dq e^{q^2/2}}\;.
\end{equation}
At the limit $\Lambda\rightarrow\infty$, $Q_\Lambda\rightarrow \infty$, $C_n(Q_\Lambda)$ is constant.
The variance of $p_N^2$ now becomes
\begin{equation}
    \text{Var}_\Lambda(p_N^2)=\langle p_N^4\rangle_\Lambda-\langle p_N^2\rangle_\Lambda^2= (m\lambda^2)^{2}\beta^{4}(C_2-C_1^2)\sim(m\lambda^2)^{2}\beta^{4} \;,
\end{equation}
where $C_2-C_1^2>0$. Then, the variance of $A$ is 
\begin{equation}
    \text{Var}_\Lambda(A)\sim \frac{(m\lambda^2)^2\beta^4}{m^2\lambda^4\beta^6}=\frac{C}{\beta^2}\;.
\end{equation}
where $C$ is a new dimensionless constant and can be absorbed by rescaling the metric. Taking the limit $\Lambda\rightarrow \infty$, we obtain 
\begin{equation}
    \lim_{\Lambda\rightarrow \infty}g^{(\Lambda)}_{\beta\beta}=\frac{C}{\beta^2}
\end{equation}
The resulting metric therefore captures the intrinsic information geometry of the parametrised family, independent of regularisation artifacts. The corresponding line element is therefore
\begin{equation}
    ds^2=\frac{d\beta^2}{\beta^2}\;.
\end{equation}
This metric is invariant under scale transformations $\beta\rightarrow a\beta$, where $a$ is scalar, reflecting the fact that only relative, rather than absolute, values of $\beta$ are statistically distinguishable. Moreover, the geodesic equation admits the solution $\beta(\mu)=\beta_0e^{\mu}$, where $\mu$ is the affine parameter. The constants $\beta_0$ is to be determined. Introducing the logarithmic coordinate $u=\ln \beta$, where $\beta >0$, the metric becomes
\begin{equation}
    ds^2=du^2\;.
\end{equation}
In this coordinate, the geometry is flat (Euclidean), but the original scale invariance is preserved through the exponential mapping between $  u  $ and $  \beta  $. 
%
\begin{figure}[h]
\centering
\includegraphics[width=1\linewidth]{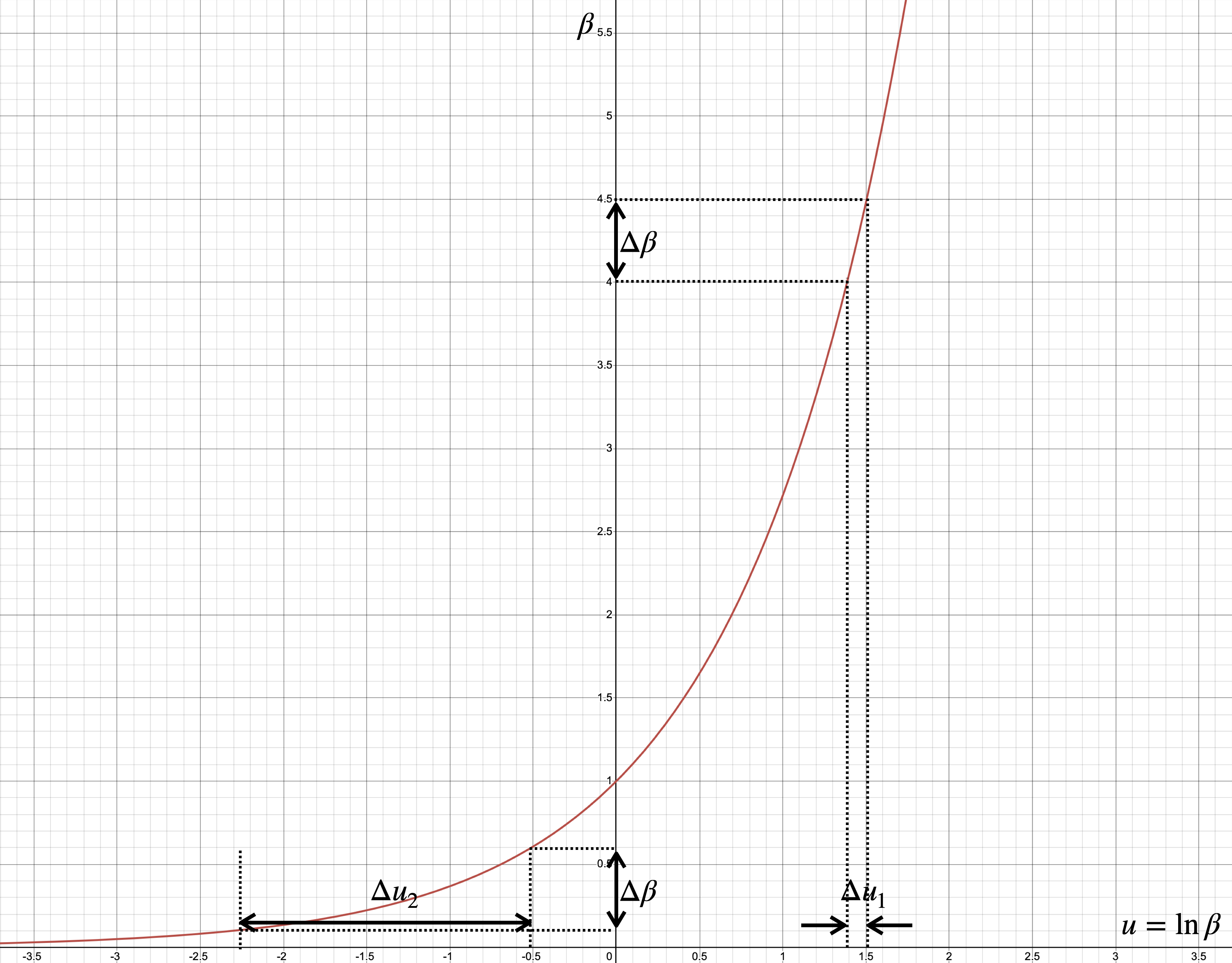}
\caption{\label{ub} Relation between $\beta$ and geodesic coordinate $u=\ln \beta$ on the Fisher-Rao statistical manifold. Equal increment in $\beta$ correspond to larger statistical distances near $\beta \rightarrow 0$, reflecting the scale-invariant geometry induced by information metric. }
\end{figure}
\\
\\
Figure \ref{ub} illustrates the statistical manifold through 2-dimensional space. 
Imagine the manifold associated with $\beta$ as a landscape (line) where the ``ground" gets more stretched out as $\beta$ approaches zero, since $1/\beta^2$ grows. In the coordinate $u$, the Fisher-Rao metric is 1, representing a ``flat" geometry where the statistical distance is simply the Euclidean distance. Together, the exponential curve, which is nothing but the geodesic solution $\beta=e^u$, where $\beta_0=1$ and $\mu=u$, explains why the landscape appears distorted near $\beta\rightarrow 0$. This logarithmic flattening highlights the intrinsic scale invariance of the statistical distinguishability among the family of multiplicative Hamiltonians.


\section{Constraints revisited: geometric origin of the relativistic Hamiltonian}
The maximum-entropy distribution $\rho(\beta)$ in section 3 is determined by three constraints: normalization, $\langle 1/\beta^2\rangle=C_1$, and $\langle \ln \beta \rangle=C_2$. While normalisation is conventional, the other two gain intrinsic justification from the Fisher–Rao geometry of the statistical manifold. Two quantities are geometrically privileged as follows
\begin{itemize}
    \item $\ln\beta$: The natural affine (geodesic) coordinate on the manifold.
    \item $1/\beta^2$: The local density of the metric $ds^2=d\beta^2/\beta^2$, which sets the scale of statistical distinguishability.
\end{itemize}
Fixing $\langle \ln \beta \rangle$ anchors the mean position along the geodesic, while fixing $\langle 1/\beta^2\rangle$ controls the average metric scale. These are exactly the constraints used. Hence they reflect the manifold's intrinsic structure, not arbitrary choices.
\\
\\
\textbf{Stability and minimality of the entropy constraints}. 
Only this specific pair produces a normalizable distribution compatible with scale invariance and yields a well-defined ensemble average leading to the relativistic form. Alternatives fail systematically:
\begin{itemize}
    \item $\langle \beta^2\rangle$ or $\langle \beta\rangle$ instead of $\langle 1/\beta^2\rangle$ breaks scale invariance.
    \item  $\langle \ln \beta\rangle$ alone leaves the average ill-defined.
    \item $\langle 1/\beta^2\rangle$ alone collapses $\rho(\beta)$ to a delta function (no averaging).
    \item Additional or substituted constraints (e.g., $\langle \beta\rangle$, higher moments) yield non-normalisable or degenerate distributions.
\end{itemize}
The chosen constraints are thus minimal and geometrically distinguished, not ad hoc. From this view, entropy maximisation in section 3 is a variational principle on the space of Hamiltonian parametrizations that respects the manifold's intrinsic geometry. The resulting $\rho(\beta)$ is the least-biased distribution compatible with the Fisher–Rao manifold's global (geodesic) and local (metric) structure.

\section{Conclusion}
In this work, we have demonstrated that the relativistic Hamiltonian emerges as an effective description from a purely statistical and information-theoretic framework, without assuming Lorentz symmetry or relativistic postulates at an early stage of the construction. The construction relies on three key elements: a multiplicative Hamiltonian structure, fluctuations of an auxiliary parameter $\beta$, and principled maximum entropy inference guided by the Fisher–Rao geometry of the associated statistical manifold.
\\
\\
The probability distribution $\rho(\beta)$ governing these fluctuations is uniquely determined by maximising Shannon entropy under scale-invariant constraints. While these constraints may initially appear formal, the Fisher–Rao analysis reveals their intrinsic meaning: $\langle \ln \beta \rangle$ fixes the mean geodesic position on the manifold, and $\langle 1/\beta^2\rangle$ sets the average information scale (metric density). Entropy maximisation thus becomes tightly coupled to the underlying geometry of Hamiltonian parameterisations. The resulting distribution, when used for ensemble averaging, yields the familiar relativistic form $\langle H \rangle=\sqrt{p^2c^2+m^2c^4}$. Remarkably, this square-root structure reflects the norm induced by statistical distinguishability in the logarithmic geometry $(ds^2=d\beta^2/\beta^2 \rightarrow \text{flat in}\;u=\ln \beta)$, rather than any imposed kinematical assumption. This insight arises already within a simple one-dimensional statistical manifold, indicating that minimal geometric complexity suffices. From a broader viewpoint, the work illustrates a concrete intersection of classical mechanics, information-theoretic inference, and statistical manifold geometry. The multiplicative Hamiltonian provides the starting point, maximum entropy supplies a least-biased inference rule under incomplete information, and Fisher–Rao geometry encodes invariant distinguishability. The relativistic dispersion relation then appears as a natural effective outcome of this interplay, rather than a fundamental axiom.
\\
\\
The present analysis is limited to single-particle kinematics and the emergence of the dispersion relation. Full aspects of special relativity, including Lorentz transformations, multi-particle interactions, and spacetime structure, lie beyond scope and are deferred to future work. Moreover, the framework developed here would offer a flexible approach that may extend to other contexts involving effective relativistic behavior, non-quadratic dispersion relations, or emergent kinematics. We hope these results could be used as a compass to explore the information geometry's role in foundations of classical and quantum contexts.

\section*{Acknowledgments}

The author would like to acknowledge the New Year holiday period, during which an unusual degree of intellectual freedom, unencumbered by immediate obligations, allowed ideas to develop organically. 

\bibliographystyle{unsrt}

\end{document}